\documentclass[superscriptaddress,showpacs,preprintnumbers,amsmath,amssymb,
nofootinbib,twocolumn,aps,prd,10pt]{revtex4-1}
\usepackage{amssymb,amsmath}
\usepackage{epsfig}
\usepackage[dvipsnames]{xcolor}
\usepackage[utf8]{inputenc}
\usepackage{stmaryrd}
\usepackage{mathrsfs}
\usepackage{mathalfa}
\usepackage[normalem]{ulem}

\newcommand{\la}{\langle}
\newcommand{\ra}{\rangle}
\newcommand{\be}{\begin{equation}}
\newcommand{\ee}{\end{equation}}
\newcommand{\ba}{\begin{eqnarray}}
\newcommand{\ea}{\end{eqnarray}}
\newcommand{\beg}{\begin{gather*}}
\newcommand{\eng}{\end{gather*}}
\newcommand{\hh}{,\hspace{0.5cm}}
\newcommand{\hhh}{,\hspace{0.2cm}}

\newcommand{\eq}[1]{(\ref{#1})}

\newcommand{\n}[1]{\label{#1}}

\newcommand{\ins}[1]{{\mbox{\tiny #1}}}
\newcommand{\ind}[1]{{\mbox{\scriptsize #1}}}
\newcommand{\inds}[1]{{\scriptscriptstyle #1}}

\def\XXint#1#2#3{{\setbox0=\hbox{$#1{#2#3}{\int}$ }
\vcenter{\hbox{$#2#3$ }}\kern-.6\wd0}}

\usepackage[makeroom]{cancel}
\usepackage[caption=false]{subfig}
\usepackage[colorlinks=true,
            citecolor=green,
            linkcolor=red,
            filecolor=cyan,
            urlcolor=magenta,
            backref=false]{hyperref}

\newcommand{\dd}{\mbox{d}}

\newcommand{\SET}{$\langle \hat{T}^{\mu\nu}(x)\rangle_\ind{ren}$}

\begin{document}

\title{Ghost-free modification of the Polyakov action and Hawking radiation}

\author{Jens Boos}
\email{boos@ualberta.ca}
\affiliation{Theoretical Physics Institute, University of Alberta, Edmonton, Alberta, Canada T6G 2E1}
\author{Valeri P. Frolov}
\email{vfrolov@ualberta.ca}
\affiliation{Theoretical Physics Institute, University of Alberta, Edmonton, Alberta, Canada T6G 2E1}
\author{Andrei Zelnikov}
\email{zelnikov@ualberta.ca}
\affiliation{Theoretical Physics Institute, University of Alberta, Edmonton, Alberta, Canada T6G 2E1}


\begin{abstract}

In this paper we discuss possible effects of non-locality in black hole spacetimes. We consider a two-dimensional theory in which the action describing matter is a ghost-free modification of the Polyakov action. For this purpose we write the Polyakov action in a local form by using an auxiliary scalar field and modify its kinetic term by including into it a non-local ghost-free form factor. We demonstrate that the effective stress-energy tensor is modified and we study its properties in a background of a two-dimensional black hole. We obtain the expression for the contribution of the ghost-free auxiliary field to the entropy of the black hole.
We also demonstrate that if the back-reaction effects are not taken into account, such a ghost-free modification of the theory does not change the energy flux of the Hawking radiation measured at infinity. We illustrate the discussed properties for black hole solution of a 2D dilaton gravity model which admits a rather complete analytical study.


\end{abstract}

\maketitle

\section{Introduction}

Non-local field theories have a long history, especially in the context of
attempts to manage the ultraviolet (UV) behavior of quantum scattering amplitudes. There is a subclass of non-local field theories that are called {\it ghost-free} (GF) theories, which have particularly nice properties. First, these ghost-free modifications of local field theories do not lead to any extra propagating degrees of freedom at tree level. As a consequence, while improving the behavior at short distances, their behavior at large scales is very similar to that of local theories. These theories have been extensively studied in a large number of publications, especially in the context of the resolution of cosmological as well as black hole singularities \cite{Tomboulis:1997gg,Biswas:2011ar,Modesto:2011kw,Biswas:2013cha,Shapiro:2015uxa,Chapiro:2019wua,Ribeiro:2018pyo,Biswas:2010zk,Biswas:2013kla,Biswas:2005qr,Edholm:2016hbt,Conroy:2015nva,Modesto:2017sdr,Buoninfante:2018xiw,Koshelev:2018hpt,Kilicarslan:2018unm,Kilicarslan:2019njc,Frolov:2015bta,Frolov:2015usa,Frolov:2015bia,Boos:2018bxf,delaCruz-Dombriz:2018aal,Buoninfante:2019swn,Kilicarslan:2018yxd,Giacchini:2018wlf}.
Scattering on a potential barrier in the framework of ghost-free theories, vacuum fluctuations and non-local footprints in observables has been analyzed in Refs.~\cite{Boos:2018kir,Boos:2019fbu,Frolov:2018bak,Buoninfante:2019teo}.
Thermal properties of ghost-free theories in flat spacetime and the superradiance effect were studied in \cite{Boos:2019zml,Frolov:2018bak}.

There is an interesting related question as to how non-locality would affect the excitation rate of an Unruh--DeWitt detector interacting with a ghost-free quantum field on the background of a Rindler spacetime or a Schwarzschild black hole. There has been some controversy in the literature on this topic, see e.g.~\cite{Nicolini:2009dr,Modesto:2017ycz,Kajuri:2017jmy,Kajuri:2018myh}. The result of this discussion can be summarized as follows: An Unruh--DeWitt detector is not sensitive to the non-locality of a ghost-free quantum field and will react exactly in the same way as if it were interacting to a local field. The explanation of this fact is quite simple. The response rate of the Unruh--DeWitt detector is described by the temporal Fourier transform of the Wightman function of the corresponding quantum field. The Wightman function satisfies a homogeneous equation and, for this reason, is the same as for the local theory. Only quantities that are described by the Feynman propagator or the retarded Green function, which satisfy \emph{inhomogeneous} equations, may be connected with non-local aspects of the field \cite{Boos:2019fbu}.

The flux of Hawking radiation of an evaporating black hole is described by the retarded propagator of the corresponding scalar field \cite{FrolovVilkovisky:1981}. Therefore it might depend on the scale of non-locality inherent to ghost-free theories. On a given background of a black hole the Hawking temperature is defined by the geometry of the black hole and it evidently depends on the surface gravity of the horizon only. The vacuum stress-energy tensor and the value of the Hawking flux at infinity in their turn depend also on the grey-body factors and on the characteristics of the quantum field. Therefore the question about the effect of non-locality on the quantum mean value of the stress-energy tensor of ghost-free fields is non-trivial.

In this paper we study quantum aspects of ghost-free theories in the strong field regime. Dealing with non-locality in this context is not a simple problem. It requires serious modifications of well-known approaches as well as the development of entirely new approaches. Our main interest in this paper is to study how a ghost-free modification of a local theory influences the stress-energy tensor of the matter field and the Hawking radiation in particular. We start with two-dimensional gravity, wherein the background geometry of a two-dimensional black hole is considered to be classical. The matter field, on the other hand, shall be assumed to be quantum.

It is well-known that in the case of a local two-dimensional conformal scalar field, a quantum mean value of its stress-energy tensor can be obtained as a variation of the Polyakov action \cite{Polyakov:1981re}, see Eq.~\eqref{a.2} further below.
This action appears as an effective action after quantization of conformal matter fields in a given background geometry and can be obtained by functional integration of the conformal anomaly \cite{Luscher:1980fr,Polyakov:1981re,Dowker:1993rt}. By construction, the Polyakov action and other effective actions are generically non-local functionals of the background fields and geometry. When the matter field is non-local even before quantization, the effective action will be non-local in any case. We do not propose a particular rigorous prescription of quantization of non-local theories. Instead we notice that, because the ghost-free modification of the matter field does not introduce any new degrees of freedom compared to the local theory, one can reasonably expect that after the quantization the effective theory also will not acquire extra poles in the propagators. Ghost-free modifications, in the class of theories considered here, contain a dimensional parameter of fundamental length $\ell$ which breaks conformal invariance of the theory. This fact can have an imprint on all quantum averages of observables.

Let us consider a class of theories
\begin{align}
S[g{}_{\mu\nu},{\hat{\psi}}] = S_\text{g}[g{}_{\mu\nu}] +  S_\ins{matter}[g{}_{\mu\nu},{\hat{\psi}}]
\end{align}
where $S_\text{g}$ is a gravitational action and $S_\text{matter}$ is the action of the quantum matter field $\hat{\psi}$ in the background of $g$. After the quantization of the matter fields $\hat{\psi}$, as well as renormalization of the coupling constants of the gravitational action, one obtains the effective action
\begin{align}
W[g{}_{\mu\nu}] = S_\text{g}[g{}_{\mu\nu}] +  W_\text{matter}[g{}_{\mu\nu}] \, .
\end{align}
Now suppose that $g{}_{\mu\nu}$ is a black hole solution of $S_\text{g}[g{}_{\mu\nu}]$. In this given background, we may ask how the ghost-free deformation of the theory, described by $W_\text{matter}$, affects the effective stress-energy tensor and the Hawking radiation of the black hole in particular.

The paper is organized as follows. In Section~\ref{sec:2} we present the standard Polyakov action in a local form by introducing an auxiliary scalar field and the effective stress-energy tensor associated with this action. We also discuss a relation between the choice of the state and zero modes of the $\Box$-operator. In Section~\ref{sec:3} we describe a ghost-free modification of the Polyakov action in the local form, obtain an expression for the effective stress-energy tensor of such a theory, and demonstrate that this tensor can be explicitly written as a sum of two terms. The first one depends on the form factor of the modified theory but is insensitive to the choice of the state. The second term, describing the dependence of the stress-energy tensor of the state, does not `feel' the presence of non-locality and coincides with the corresponding expression for the original local (non-modified) theory. Non-local contribution to the entropy of a 2D black hole is discussed in Section~\ref{sec:4}. In Section~\ref{sec:5} we demonstrate that for a fixed background of a 2D black hole the non-local modification of the theory does not change the energy flux of Hawking radiation at spatial infinity. In Section~\ref{sec:6} we analyze non-local effects for a special case of 2D black hole model connected to string theory and obtain explicit expressions for the components of the effective stress-energy tensor as well as contributions to the quantum corrections of the black hole entropy. The last Section~\ref{sec:7} contains a brief summary and discussion of the obtained results. Useful formulas for the 2D geometry of a static lack hole are collected in Appendix~\ref{app}.

\section{2D conformal anomaly and Polyakov action}
\label{sec:2}

Let us consider a two-dimensional spacetime with a metric $g_{\mu\nu}$, and let $\hat{\psi}$ be a conformally invariant quantum field in this metric. Then, as it is well known, the quantum average of the trace of the stress-energy tensor for such a field has a following universal form:
\be\n{a.1}
\la \hat{T}^{\mu\nu}\ra g_{\mu\nu}=2b R\, .
\ee
Here, $R$ is the Ricci scalar for the metric $g_{\mu\nu}$. The dimensionless coefficient $b$ depends on the nature of the quantum field. For a conformal massless scalar field $b=1/(48\pi)$. Polyakov \cite{Polyakov:1981re} demonstrated that the expression for the trace anomaly \eqref{a.1} can be obtained by variation of the following non-local effective action:\footnote{We use Misner--Thorne--Wheeler sign conventions for the definition of the Riemann tensor and the signature $(-,+)$ \cite{Misner:1974qy}.}
\be\n{a.2}
W_\ins{Pol}[g_{\mu\nu}]=-{b\over 2}\int \dd^2 x\, \sqrt{-g} \,R \, {1\over \Box} R\, .
\ee
One obtains
\be\n{a.3}
T=T^{\mu\nu}g_{\mu\nu}={2\over \sqrt{-g}}{\delta W_\ins{Pol}\over \delta g_{\mu\nu}} g_{\mu\nu} =2 bR\, .
\ee

The Polyakov action (\ref{a.2}) can be identically rewritten in a local form by introducing an auxiliary field $\varphi$. To that end, let us consider the action
\be\n{a.4}
W_\ins{Pol}[g{}_{\mu\nu},\varphi]=b \int \dd^2 x\ \sqrt{-g}\left[{1\over 2} \varphi \,\Box \varphi -R\varphi\right]\, .
\ee
The scalar curvature here plays the role of a source for the field $\varphi$.
The variation of this action with respect to $\varphi$ gives
\be\n{a.5}
\Box \varphi=R\, .
\ee

By substituting this relation into (\ref{a.4}) one returns to the Polyakov action (\ref{a.2}). It should be emphasized that the effective action (\ref{a.4}) depends on the classical fields $g_{\mu\nu}$ and $\varphi$. It correctly reproduces the conformal trace anomaly, but itself it is \emph{not} conformally invariant.

The effective action can be used to calculate not only the trace, but all components of the effective stress-energy tensor \cite{FrolovVilkovisky:1981}
\be\begin{split}\n{a.6}
&T^{\mu\nu}={2\over \sqrt{-g}} {\delta W_\ins{Pol}\over \delta g_{\mu\nu}}\\
&=b\Big[\varphi^{;\mu}\varphi^{;\nu}-2 \varphi^{;\mu\nu}
-g^{\mu\nu}\Big( {1\over 2}\varphi^{;\alpha}\varphi_{;\alpha}-2\Box\varphi  \Big)\Big] \, .
\end{split}\ee
In this expression we understand
\be\n{a.7}
\varphi={1\over \Box}R\, .
\ee
The tensor \eqref{a.6} is conserved, $T^{\mu\alpha}{}_{;\alpha}=0$, and its trace reproduces \eq{a.3}.
The expression ${1/\Box}$ should be understood as a corresponding Green function of the $\Box$-operator.
In the application to spacetimes with Lorentzian signature of the metric, this Green function is to respect the initial conditions of the problem under consideration. If there are no incoming fluxes one should use the retarded Green function. This choice corresponds to the calculation of the $\langle \text{in}|\hat{T}^{\mu\nu} | \text{in} \rangle$  quantum mean value of the stress-energy tensor operator of the conformal quantum field $\hat{\psi}$.\footnote{For more details see the discussion in Ref.~\cite{FrolovVilkovisky:1981}.}

Let us now turn to black holes. A static two-dimensional metric can be written in the form
\be\n{a.8}
\dd s^2=-f\, \dd t^2+{\dd r^2\over f} \, ,
\ee
where $f=-\xi^{\mu} \xi_{\mu}$ and $\xi_{\mu}$ is the Killing vector (see Appendix~\ref{app:2d}). A solution of equation (\ref{a.5}) can be written as a sum
\be\n{a.9}
\varphi=\Phi_0+\chi \, ,
\ee
where
\be
\Phi_0=-\ln f
\ee
is a solution of the inhomogeneous equation and $\chi$ is a solution of the homogeneous equation $\Box\chi=0$. In other words, $\chi$ is a solution constructed from zero modes of the $\Box$-operator. We are looking for solutions that generate a stationary stress-energy tensor (\ref{a.6}). Such zero modes can be written in the following form (see Appendix~\ref{app}):
\be\n{chi}
\chi= w t+k r_*\, .
\ee
Here, $w$ and $k$ are two arbitrary constants and $r_*$ is a tortoise coordinate,
\be
r_*=\int {\dd r\over f}\, .
\ee

Substituting (\ref{a.9}) into (\ref{a.6}) one obtains
\be
T^{\mu\nu}=  T_\inds{(\Phi_0)}^{\mu\nu} + T_\inds{(\chi)}^{\mu\nu} .
\ee
In the above, $T_\inds{(\Phi_0)}^{\mu\nu}$ denotes the contribution of the $\Phi_0$-term and $T_\inds{(\chi)}^{\mu\nu}$ corresponds to the contribution of zero modes, respectively. The first term is
\be
T_\inds{(\Phi_0)}^{\mu\nu}=b\Big[\Phi_0^{;\mu}\Phi_0^{;\nu}-2\Phi_0^{;\mu\nu}-g^{\mu\nu}\Big( {1\over 2}\Phi_0^{;\alpha}\Phi_0{}_{;\alpha}-2\Box\Phi_0  \Big)\Big] .
\ee
Simple calculations allow one to write the following explicit expression
\be
T_\inds{(\Phi_0)}{}^{\mu}_{\nu}= b \left(
   \begin{array}{cc}
       -2f''+{f'^2\over 2 f} & 0 \\
       0 & -{f'^2\over 2 f} \\
     \end{array}
   \right).
\ee
The zero-mode contribution reads
\be\label{TchiLocal}
T_\inds{(\chi)}{}^{\mu}_{\nu}= b \left(
   \begin{array}{cc}
       -{k^2+w^2\over 2 f} & wk \\
       -{wk\over f^2} & {k^2+w^2\over 2 f} \\
     \end{array}
   \right) \, .
\ee

Different choices for the constants $w$ and $k$ correspond to different states of the quantum field:
When either $w$ or $k$ vanishes the non-diagonal elements of the stress-energy tensor vanish as well, that is, for such a choice of the state there are no fluxes.
The choice $w=0$, $k=2\kappa$, where $\kappa$ is the surface gravity
\be
\kappa={1\over 2}f'|_{r=r_g} ,
\ee
corresponds to the Hartle--Hawking state. Lastly, $w=\kappa$, $k=-\kappa$ defines the Unruh vacuum state. For details see Appendix~\ref{app:2d}.


\section{Ghost-free modification of the Polyakov action}
\label{sec:3}
\subsection{Action}

Let us now study a non-local modification of the action (\ref{a.4}) by substituting instead of the $\Box$-operator its ghost-free version.
To that end, let us consider a non-minimally coupled ghost-free real scalar field in two dimensions:
\begin{align}
\label{eq:action}
W_\ins{GF}[g{}_{\mu\nu}, \varphi] = \frac{1}{48\pi} \int\dd^2 x \sqrt{-g} \left[ \frac12 \varphi A \varphi - R\varphi \right] \, ,
\end{align}
where the background is given by  metric $g$ and the operator $A$ is
\begin{align}\n{P.1}
A = \Box \, e^{P(\Box)} \, , \quad P(z) = (-\ell^2 z)^N \,  .
\end{align}
Here $N$ is a positive integer number. We refer to this class of theories at to $\mathrm{GF_N}$ \cite{Frolov:2016xhq}.
The scalar field equation is
\begin{align}\n{P.2}
A \varphi = R \, ,
\end{align}
where the Ricci scalar acts as a source for the field $\varphi$. Integrating out the scalars from \eqref{eq:action} one obtains the action
\begin{align}
\begin{split}
\label{eq:gf-polyakov}
W_\ins{GF}[g{}_{\mu\nu}] &= -\frac{1}{96\pi}\int\dd^2 x \sqrt{-g} R A^{-1} R \\
&= -\frac{1}{96\pi}\int\dd^2 x \sqrt{-g} R \frac{e^{-(-\ell^2\Box)^N}}{\Box} R \, .
\end{split}
\end{align}
In the limit $\ell\rightarrow 0$ it corresponds to the Polyakov action.
For non-vanishing parameter of the non-locality $\ell$  the above action \eqref{eq:gf-polyakov}
is a \emph{ghost-free deformation} of the Polyakov action. Our aim is to analyze how this modification affects physical observables.

For the case of $\mathrm{GF_1}$ theory the action (\ref{eq:gf-polyakov}) can be written in the form
\ba
&&W_\ins{GF}=W_\ins{Pol}+W_{\ell}\, ,\nonumber\\
&&W_{\ell}=\int\limits_0^{\ell^2}\dd s \, \widetilde{W}[s]\, ,\n{WWW}\\
&& \widetilde{W}[s]=-\frac{1}{96\pi} \int\dd^2 x \sqrt{-g} \, R\, e^{s\Box} R \, .\nonumber
\ea
In the above, $W_\ins{Pol}$ denotes the Polyakov action (\ref{a.2}) and the term $W_{\ell}$ describes its ghost-free modification. In the limit $\ell \rightarrow 0$ one has $W_\ell = 0$ and one arrives back at the standard theory without ghost-free modifications.

\subsection{Trace of the effective stress-energy tensor}
\label{TRACE}
We derive here an expression for the trace of the stress-energy tensor for the action (\ref{eq:gf-polyakov}). The expression for the complete stress-energy tensor will be given in the next subsection.

For the calculation of the trace we write a two-dimensional metric in the conformal gauge,
\begin{align}
g{}_{\mu\nu} = e^{2\sigma}\eta{}_{\mu\nu} \, , \quad \sqrt{-g} = e^{2\sigma} \, .
\end{align}
Using this representation one obtains
\begin{align}
\Box = e^{-2\sigma}\overline{\Box} \, , \quad R = -2\Box\sigma \, ,
\end{align}
where $\overline{\Box}$ denotes the flat d'Alembertian. The ghost-free Polyakov action \eq{eq:gf-polyakov} then takes the form
\begin{align}
W_\ins{GF}[\sigma] &= -\frac{1}{24\pi} \int\dd^2 x \,e^{2\sigma} \sigma  e^{-(-\ell^2\Box)^N}\Box \sigma \, .
\end{align}
The trace of the energy-momentum tensor for the scalar field can be obtained via
\begin{align}\n{GFT}
T = g{}_{\mu\nu}T{}^{\mu\nu} = \frac{2g{}_{\mu\nu}}{\sqrt{-g}} \frac{\delta W_\ins{GF}}{\delta g{}_{\mu\nu}} = e^{-2\sigma} \frac{\delta W_\ins{GF}}{\delta \sigma} \, .
\end{align}
Note that in the conformal gauge
\begin{align}
\delta(\sqrt{-g}\Box) = \delta\overline{\Box} = 0 \, , \quad \delta\Box = -2\delta\sigma\Box \, .
\end{align}

The only term in Eq.~\eqref{GFT} which requires new calculational techniques is $e^{-P(\Box)}$. The variation of the exponent of an operator can be performed using the following relation \cite{Snider:1964,Wilcox:1967zz}:
\begin{align}\label{variation}
\delta \left( e^{\hat{B}} \right) = \int\limits_0^1 \dd \xi e^{(1-\xi)\hat{B}} \left( \delta\hat{B} \right) e^{\xi\hat{B}} \, ,
\end{align}
which is applicable to variation of an exponent of any self-adjoint operator $\hat{B}$. This relation allows one to obtain the expression for the trace for an arbitrary $\mathrm{GF_N}$ model. Here we present the corresponding result for the simplest case of $\mathrm{GF_1}$ theory. We obtain
\begin{align}
\begin{split}
\label{eq:trace-gf1}
T &= \frac{1}{24\pi} e^{\ell^2\Box} R \\
&+\frac{\ell^2}{48\pi}\int\limits_0^1\dd\xi \left[ e^{(1-\xi)\ell^2\Box} R \right] \left[ e^{\xi \ell^2 \Box} R \right]  .
\end{split}
\end{align}
For $\ell=0$ this expression correctly reproduces the trace anomaly of the Polyakov action $T=\frac{1}{24\pi} R$.


\subsection{The effective stress-energy tensor}

Variation of the action \eq{eq:action}--\eq{eq:gf-polyakov} over the metric gives the effective stress-energy tensor. For example, in the case of $\mathrm{GF_1}$ theory one has $B=P(\Box)=-\ell^2\Box$ and the effective stress-energy tensor reads
\be\begin{split}\n{Tmunu}
T^{\mu\nu}={1\over 48\pi}&\Big[
\varphi^{;\mu}(e^{-\ell^2\Box}\varphi)^{;\nu}-{1\over 2}g^{\mu\nu}\varphi^{;\alpha}(e^{-\ell^2\Box}\varphi)_{;\alpha}\\
&-2 \varphi^{;\mu\nu}
+2g^{\mu\nu}\Box\varphi \Big] \\
-{\ell^2\over 48\pi}&\int\limits_0^1\dd\xi\Big\{
\Big(e^{-(1-\xi)\ell^2\Box}\Box\varphi\Big)^{;\mu} \Big(e^{-\xi \ell^2\Box}\varphi\Big)^{;\nu}\\
&-{1\over 2}g^{\mu\nu}\Big(e^{-(1-\xi)\ell^2\Box}\Box\varphi\Big)^{;\alpha} \Big(e^{-\xi \ell^2\Box}\varphi\Big)_{;\alpha}\\
&-{1\over 2}g^{\mu\nu}\Big(e^{-(1-\xi)\ell^2\Box}\Box\varphi\Big) \Big(e^{-\xi \ell^2\Box}\Box\varphi\Big)
\Big\}.
\end{split}\ee
Here, after the variation, one can apply the field equation (\ref{P.2}) and write
\be\label{varphi}
\varphi=A^{-1}R={e^{\ell^2\Box}\over\Box}R.
\ee
These formulae generalize \eq{a.6}--\eq{a.7} to the non-local $\mathrm{GF_1}$ theory. Similarly, one can easily derive the stress-energy tensor for arbitrary $\mathrm{GF}_N$ theory for other values of $N$. It is straightforward to verify that by taking trace of \eq{Tmunu} one correctly reproduces (\ref{eq:trace-gf1}).

\subsection{State dependence}

A solution (\ref{varphi}) of the equation (\ref{P.2}) for $GF_1$ model can be written as a sum
\be\n{Phi}
\varphi=\Phi+\chi\, ,
\ee
were $\Phi$ is a solution of the inhomogeneous equation (\ref{P.2})
\be\label{PhiPhi0}
\Phi=e^{\ell^2\Box}\Phi_0 \, , \quad \Phi_0=-\ln f\, ,
\ee
and $\chi$ is a zero mode of the operator $A$. Since the form factor $e^{\ell^2\Box}$ calculated for an on-shell solution is equal to 1, zero modes of the operator $A$  are identical to zero modes $\chi$ of the $\Box$-operator
\be
\Box\chi=0\, .
\ee
This is a property intrinsic to ghost-free theories and \emph{not} present in generic higher-derivative theories. Using these results one can show that  \eq{Tmunu} splits into two terms,
\be
T^{\mu\nu}=T_\inds{(\Phi)}^{\mu\nu}+T_\inds{(\chi)}^{\mu\nu}\, .
\ee

The first term is given by the same formula as \eq{Tmunu} where $\varphi$ is replaced by $\Phi$,
\begin{align}\begin{split}\n{TPhi}
T_\inds{(\Phi)}^{\mu\nu}={1\over 48\pi}&\Big[
\Phi^{;\mu}(e^{-\ell^2\Box}\Phi)^{;\nu}-{1\over 2}g^{\mu\nu}\Phi^{;\alpha}(e^{-\ell^2\Box}\Phi)_{;\alpha}\\
&-2 \Phi^{;\mu\nu}
+2g^{\mu\nu}\Box\Phi \Big] \\
-{\ell^2\over 48\pi}&\int\limits_0^1\dd\xi\Big\{
\Big(e^{-(1-\xi)\ell^2\Box}\Box\Phi\Big)^{;\mu} \Big(e^{-\xi \ell^2\Box}\Phi\Big)^{;\nu}\\
&-{1\over 2}g^{\mu\nu}\Big(e^{-(1-\xi)\ell^2\Box}\Box\Phi\Big)^{;\alpha} \Big(e^{-\xi \ell^2\Box}\Phi\Big)_{;\alpha}\\
&-{1\over 2}g^{\mu\nu}\Big(e^{-(1-\xi)\ell^2\Box}\Box\Phi\Big) \Big(e^{-\xi \ell^2\Box}\Box\Phi\Big)
\Big\}.
\end{split}
\end{align}
The zero-mode dependent term $T_\inds{(\chi)}^{\mu\nu}$ reads
\be\label{Tchi}
\begin{split}
T_\inds{(\chi)}^{\mu\nu}&={1\over 48\pi}\Big[
\chi^{;\mu}\chi^{;\nu}
-{1\over 2}g^{\mu\nu}\chi^{;\alpha}\chi_{;\alpha}-2 \chi^{;\mu\nu} \Big]\\
&+{1\over 48\pi}\Big[
\Phi_0^{;\mu}\chi^{;\nu}+\chi^{;\mu}\Phi_0^{;\nu}-g^{\mu\nu}\,\chi_{;\alpha}\Phi_{0}^{\alpha}\Big]\, .
\end{split}\ee
Using the field equation $\Box\chi=0$ one can check that its trace vanishes, $T_\inds{(\chi)}=0$. The component $T_\inds{(\chi)}^{\mu\nu}$ does not depend on the parameter $\ell$ and hence it coincides with the corresponding expression for the Polyakov action discussed earlier. One can see that the non-diagonal components of the effective stress-energy tensor describing the fluxes are given by the state dependent term $T_\inds{(\chi)}^{\mu\nu}$. Therefore one can conclude that, if the back-reaction of the effective stress-energy tensor on the metric is neglected, the Hawking flux of energy at infinity does not feel the effects of non-locality. At the same time the diagonal part $T_\inds{(\Phi)}^{\mu\nu}$ of the stress-energy tensor depends on the non-locality parameter $\ell$ and its back-reaction on the metric modifies the parameters of the black hole.


\section{Black hole entropy}
\label{sec:4}

The representation \eq{eq:action} of the ghost-free action is useful for determining the contribution of ghost-free fields to the quantum corrections of black hole entropy. As it has been proved by Myers \cite{Myers:1994sg}, the Noether charge technique proposed by Wald \cite{Wald:1993nt} can be successfully applied to non-local theories as well. The purely gravitational part of the action $S_g[g_{\mu\nu}]$ is local and obviously leads to a standard Wald contribution to the entropy of the black hole. In what follows, we shall be interested in the part stemming from the ghost-free action \eq{eq:gf-polyakov}. In this case, employing the local representation \eq{eq:action}, one can easily compute the ghost-free contribution to the entropy\footnote{Here we use the letter S for the entropy, as it is traditionally accepted, although previously we used the same symbol to denote classical actions. We hope that it will not lead to confusion.}
\be\label{SGF1}
S_\ins{GF}={1\over 12}\varphi\big|_{r=r_g},
\ee
where $\varphi$ is given by \eq{varphi}. For $\mathrm{GF_1}$ theory we obtain
\be\label{SGF2}
S_\ins{GF}={1\over 12} {e^{\ell^2\Box}\over\Box}R \Big|_{r=r_g}.
\ee
In the limit $\ell\rightarrow 0$ the ghost-free action $W_\ins{GF}$ reduces to the Polyakov action $W_\ins{Pol}$ and one reproduces its standard contribution $S_\ins{Pol}$ to the black hole entropy,
\begin{align}\label{SPol}
S_\ins{Pol}={1\over 12} {1\over\Box}R \Big|_{r=r_g}.
\end{align}

Note that $\varphi$ and the propagator $1/\Box$ entering \eq{SGF2}--\eq{SPol} depend on the choice of the state, which in turn is reflected in the proper boundary conditions for the Green function. For every state these boundary conditions can be satisfied by adding zero-modes \eq{chi}. For the Hartle--Hawking vacuum the auxiliary field $\varphi$ is finite on the bifurcation point of horizons and takes the form
\be\label{c}
\varphi=\Phi+2\kappa r_*+c \, ,
\ee
where, $\Phi$ is given by \eq{PhiPhi0} and c is a constant. Similarly, in the case of the Polyakov action one gets
\be
\varphi=\Phi_0+2\kappa r_*+c \hh  \Phi_0=-\ln f \, .
\ee
Because zero modes are the same for the ghost-free and Polyakov models,
the constant $c$ here is the same as in \eq{c}. One can fix this constant by considering the pure Polyakov model, wherein it is defined by the boundary conditions and a proper gauge fixing for the conformal metric \cite{Myers:1994sg}. The difference of the entropies in these two models, $\Delta S=S_\ins{GF}-S_\ins{Pol}=(\Phi+\ln f)/12$, is finite, uniquely defined, and does not depend on the state.

\section{Hawking flux}
\label{sec:5}

As we already mentioned, the analysis of the effective stress-energy tensor allows one to conclude that the ghost-free modification of the Polyakov action does not affect the fluxes as measured at infinity. In this section we re-derive this result by using the Christensen--Fulling representation for a general stationary conserved stress-energy tensor in two-dimensional static spacetimes \cite{Christensen:1977jc}.
As is well known, in two dimensions the Hawking flux can be evaluated if a trace of the stress tensor is given. In particular, in the metric of the form (\ref{a.8}) the conservation of the stationary energy-momentum tensor gives
\begin{align}
\label{eq:tmunu-cons}
\partial_r T{}^r_t = 0 \, , \quad \partial_r( f T{}^r_r ) = \frac12 f' T{}^\alpha_\alpha \, .
\end{align}
The Hawking flux at infinity is given by \cite{Christensen:1977jc}
\begin{align}
{\dd E\over \dd t} = \frac12 \int\limits_{r_g}^\infty \dd r f'(r) T{}^\alpha_\alpha(r) \, ,
\end{align}
where $r=r_g$ corresponds to the black hole horizon such that $f(r_g) = (\nabla r)^2|_{r=r_g} = 0$. From \eqref{eq:tmunu-cons} it is clear that the Hawking flux at infinity picks up a contribution from the horizon, $f(r_g) T^r_r(r_g)$. Consequently there will be no contribution to the Hawking flux from any $T^r_r$ that is finite at the horizon.

Using the representation
$W_\ins{GF}=W_\ins{Pol}+W_{\ell}$, see \eqref{WWW}, one can write the following expressions for the trace $T$ of the effective action and its $(r,r)$ component, $T_r^r$:
\be
T = T_\ins{(Pol)} + \int\limits_0^s \dd{s} \, \widetilde{T} \, , \quad T^r_r = T_\ins{(Pol)}{}^r_r  + \int\limits_0^s \dd{s} \, \widetilde{T}^r_r \, .
\ee
Here
\be\n{TVAR}
\widetilde{T}^{\mu\nu}={2\over \sqrt{-g}} {\delta \widetilde{W}[s]\over \delta g_{\mu\nu}}
\ee
and $\widetilde{W}[s]$ is defined by Eq.~\ref{WWW}. The trace $\widetilde{T} $ can be easily found by using the method explained in subsection~\ref{TRACE}.

Let us now explain how the components of $\widetilde{T}^{\mu\nu}$ can be determined. We start with a general expression for the variation of the action $\widetilde{W}[s]$
\be \n{DW}
\delta\widetilde{W}[s]=\int \dd^2x \sqrt{-g}\, \widetilde{T}^{\mu\nu} \delta g_{\mu\nu}\, .
\ee
After the variation is performed and $\widetilde{T}^{\mu\nu}$ is obtained in an arbitrary metric, let us substitute into this expression the static metric (\ref{a.8}). We now can consider special (static) variations of the metric
\be\n{DF}
\delta({\dd s^2})=-\left[ \dd t^2+{\dd r^2\over f^2}\right] \delta f\, .
\ee
Note that the integrand in (\ref{DW}) does not depend on time. As the result of variation one obtains
\be
2f\frac{\delta \widetilde{W}[s]}{\delta f} =  \widetilde{T}^r_r -  \widetilde{T}^t_t \,  .
\ee
Using these results one finds the following expressions for $\widetilde{T} $ and $\widetilde{T}^r_r$
\begin{align}
\widetilde{T} &= \frac{1}{48\pi} \Big[ 2f e^{s\Box} R'' + 2f'\left( e^{s\Box} R \right)' + R e^{s\Box} R \nonumber \\
&\hspace{30pt} + sf \int\limits_0^1 \dd \xi \left( e^{(1-\xi)s\Box}R \right)' \left( e^{\xi s \Box} R \right)' \Big] \, , \\
\widetilde{T}^r_r &= \frac{1}{96\pi}\Big[ 4f \left(e^{s\Box}R \right)'' + 2f'\left( e^{s\Box R}\right)' + R e^{s\Box}R  \nonumber \\
&\hspace{30pt} + s \partial_r \int\limits_0^1 \dd \xi f \left( e^{(1-\xi)s\Box}R \right) \left( e^{\xi s\Box}R \right)' \Big] \, ,
\end{align}
where $(\dots)' = \partial_r(\dots)$. Let us consider these two expressions in two regimes: at the horizon where $f(r_g)=0$, and at spatial infinity where $f=1$ and $R=0$.

At spatial infinity, both $\widetilde{T}$ and $\widetilde{T}^r_r$ vanish. At the horizon they are regular and finite. Note, however, that this regularity of $\widetilde{T}^r_r$ at the horizon implies
\begin{align}
\lim\limits_{r\rightarrow r_g} f \widetilde{T}^r_r = 0 \, .
\end{align}
Then the conservation of energy momentum \eqref{eq:tmunu-cons} implies that GF modification of the Polyakov action cannot affect the total flux of Hawking radiation at spatial infinity.

\section{Example: 2D dilaton black hole}\label{Sec6}
\label{sec:6}

\subsection{Action and solutions}

In order to illustrate the effects of non-locality on the properties of black hole we consider a 2D theory described by the effective action
\begin{align}\label{2BH}
S[g{}_{\mu\nu}] = {1\over 2}\int \dd^2x \sqrt{-g}\, e^{-2\phi}[R+4(\nabla\phi)^2+4\lambda^2] \, .
\end{align}
Here $R$ is the curvature of the 2D spacetime,  $\phi$ is a dilaton field and $\lambda$ is a constant. This action arises in string theory \cite{Fradkin:1985ys,Callan:1985ia}. Its 2D black hole solutions were studied in Refs.~\cite{Witten:1991yr,Mandal:1991tz,McGuigan:1991qp,Frolov:1992xx}.\footnote{Solutions for the action (\ref{2BH}) with conformal classical and quantum matter (the so called CGHS model) have been discussed in \cite{Callan:1992rs}. For a review see Refs.~\cite{Grumiller:2002nm,Fabbri:2005mw}.}

A static black hole solution can be written as
\be
f=1-{M\over\lambda}e^{-2\lambda r} \hh \phi=-\lambda r\,.
\ee
Here, $M$ is the mass parameter of this black hole solution. The horizon is located at $r=r_g$ with
\be\label{rg}
r_g={1\over 2\lambda}\ln{M\over \lambda}\,.
\ee
The constant $\lambda$ determines a scale. It is convenient to use  dimensionless coordinates  $(\tau, x)$ defined as
\be
\tau=2\lambda t\hh x=2\lambda(r-r_g) \, ,
\ee
and write the `physical' metric $\dd\bar{s}^2$  in the form
\begin{align}
\label{eq:cghs}
\dd\bar{s}^2 = {1\over 4\lambda^2}\,  \dd s^2\hhh \dd s^2=-f\,  \dd\tau^2 + \frac{\dd x^2}{f} \hhh f = 1-e^{-x} \, .
\end{align}
In what follows we perform our calculations in the dimensionless metric $\dd s^2$ using the dimensionless coordinates $(\tau, x)$ and only at the very end restore the dimensionality of the corresponding objects. For example, the surface gravity in the physical metric is $\bar{\kappa}=(\dd f/\dd r)\big|_{r=r_g}/2=\lambda$ while in the dimensionless one it is $\kappa=1/2$.

The dimensionless Ricci curvature of the black hole is
\begin{align}
R = e^{-x} \, .
\end{align}
In what follows it will be convenient to use the dimensionless curvature $R$ instead of the coordinate $x$. In these curvature coordinates the metric (\ref{eq:cghs}) takes the form
\be \n{MR}
\dd s^2=-(1-R) \dd\tau^2+{\dd R^2\over R^2 (1-R)}\, .
\ee

\subsection{Spectral representation}
\label{sec:spectral}

Our formulae describing the contribution of non-locality to the stress-energy tensor and black hole entropy contain the quantity $F(s,R) = e^{s\Box} R$ and other functions similar to it. Let us calculate this object. This function  $F(s,R)$ obeys the following equation
\begin{align}\n{FSR}
\left(\partial_s - \Box \right)F(s,R) = 0 \, , \quad F(0,R) = R \, .
\end{align}
The second equality plays the role of an initial condition. In our calculations we shall use the curvature coordinates $(\tau,R)$ in which the $\Box$-operator takes the form
\begin{align}\n{BOX}
\Box &= R\ \partial_R\left[(1-R)\ R\ \partial_R \right]\, .
\end{align}

Let us consider the following eigenvalue problem
\be \n{EVP}
\Box \Psi(R) = \lambda \Psi(R) \, ,
\ee
and require that a real eigenfunction $\Psi(R)$ is finite both at the horizon and at infinity. It is easy to show that at the horizon, $R=1$, a general solution of (\ref{EVP}) has the following asymptotics:
\be \n{BC_1}
\Psi(R)\sim a_{-1}\ln(1-R)+a_0+\ldots \, ,
\ee
At infinity its asymptotics are
\be \n{BC_2}
\Psi(R)\sim a_{+} R^{\sqrt{\lambda}} +a_{-} R^{-\sqrt{\lambda}}\, .
\ee

Let us show that for a positive value of $\lambda$ it is impossible to satisfy simultaneously the condition of the finiteness of $\Psi$ on the horizon and at infinity. Let us denote
\be
R_*=-\ln\left({R\over 1-R}\right) \, .
\ee
Then, Eq.~(\ref{EVP}) can be written in the form
\be
{\dd^2 \Psi\over \dd R_*^2}=\lambda (1-R)\Psi\, .
\ee
The coordinate $R_*$ monotonically increases from $-\infty$ at the horizon to $\infty$ at the infinity
The finiteness at the horizon implies $a_{-1}=0$, so that
\be
\Psi\Big|_{R_*=-\infty}=a_0\hh {\dd\Psi\over \dd R_*}\Big|_{R_*=-\infty}=0\, .
\ee
The constant $a_0$ depends on the normalization of $\Psi$ and we can always choose it to be positive. Then the relation
\be
{\dd \Psi\over \dd R_*}=\lambda \int\limits_{-\infty}^{R_*}(1-R)\Psi \dd R_*=\lambda \int\limits_0^R {\dd R\over R}\Psi
\ee
implies that $\dd \Psi/ \dd R_*$ is a positive growing function of $R_*$. Hence $\Psi$ grows at infinity and the second boundary condition (\ref{BC_2}) cannot be satisfied. Thus $\lambda\le 0$. We denote $\lambda=-p^2$ and write Eq.~(\ref{EVP}) in the form
\be\n{PPSI}
\Box \Psi_p(R) = -p^2 \Psi_p(R) \,  .
\ee
For real $p$ both asymptotics $R^{\pm i p}$ remain finite at infinity (at $R\to 0$). This implies that the corresponding eigenvalue problem (\ref{EVP}) has a \emph{continuous} spectrum.

Using the eigenfunctions $\Psi_p(R)$ one can write a solution of (\ref{FSR}) in the form
\begin{align}
\tilde{F}(s,R) = \int \dd p \, \rho_p \, e^{-p^2 s} \Psi_p(R) \, .
\end{align}
Here, the spectral density factor $\rho_p$ is to be determined by the boundary condition $F(0,R) = R$.

Eigenfunctions $\Psi_p(R)$ can be found in an explicit form. For this purpose let us notice that a complex function $Z_p(R)$,
\begin{align}
Z_p(R) &= R^{ip} {}_2 F_1\left( ip, ip+1; 2ip+1; R \right) \, , \quad p \in \mathbb{R} \, ,
\end{align}
is a solution of Eq.~(\ref{PPSI}). It is easy to see that
\be
\bar{Z}_p(R)=Z_{-p}(R)\, .
\ee
Real solutions can be written in the form $\Re[{Z}_p(R)]=1/2[Z_p(R)+Z_{-p}(R)]$ and $\Im[{Z}_p(R)]=(1/2i)[Z_p(R)-Z_{-p}(R)]$. Therefore, if one does not impose the requirement that a solution is finite at the horizon for a given eigenvalue $p^2$, there exist \emph{two} real solutions. These functions can be used for the construction time-dependent propagating modes of the $\Box$-operator. In the present case, which relies on static modes, one needs to impose the condition of the finiteness the horizon.

Expanding $Z_p(R)$ close to $R=1$ one finds
\begin{align}
Z_p(R) &\approx b_p + c_p \log(1-R) + \mathcal{O}\left(1-R\right) \, , \\
b_p &= - \frac{4^{ip}\Gamma\left(ip+\tfrac12\right)}{p\sqrt{\pi}\Gamma(ip)}\big[ -i + 2p\gamma + 2p\psi(ip) \big] \, , \\
c_p &= - \frac{4^{ip}\Gamma\left(ip+\tfrac12\right)}{\sqrt{\pi}\Gamma(ip)} \, .
\end{align}
Here $\psi(ip)$ is the digamma function.
A real-valued solution that is finite both at the horizon ($R=1$) and at infinity ($R=0$), where it is oscillating, is then given by $(p\ge 0$)
\begin{align}
\Psi_p(R) = f_p \Big[ \Re(c_p)\Im(Z_p(R)) - \Im(c_p)\Re(Z_p(R)) \Big] \, .
\end{align}
For a given value $p\ge 0$, the above procedure reduces the number of solutions, that are real and finite at the horizon, down to one. Incidentally, this solution is similar to a standing wave.

\subsection{Orthogonality and normalization of the eigenfunctions}

The Wronskian of two eigenfunctions $\Psi_p$ and  $\Psi_q$ is
\begin{align}
W[\Psi_p, \Psi_q] = R(1-R)\Big[ \Psi_p(R) \overset{\leftrightarrow}{\partial}_R \Psi_q(R) \Big] \, .
\end{align}
Since solutions are finite at spatial infinity ($R=0$) as well as at the horizon ($R=1$), the Wronskian vanishes at these points. Then one obtains
\begin{align}
\begin{split}
&0= \int\limits_0^1 \dd R \, \partial_R W[\Psi_p, \Psi_q]
= (q^2-p^2) \la \Psi_p, \Psi_q\ra \, ,\n{ORT}\\
 &\la \Psi_p, \Psi_q\ra=\int\limits_0^1 \frac{\dd R}{R} \Psi_p(R) \Psi_q(R) \, .
\end{split}
\end{align}
The first equality follows because the solutions are finite at the horizon ($R=1$) and at infinity ($R=0$). The last line determines a scalar product in a space of solutions of the equation (\ref{PPSI}). The relation (\ref{ORT})  shows that eigenfunctions with different eigenvalues are orthogonal in this scalar product and these eigenfunctions with a proper choice of the normalization constant satisfy the following relation\footnote{Note that the functions $Z_p$ that diverge logarithmically at $R=1$ do \emph{not} satisfy the above orthogonality properties.}
\begin{align}
\int\limits_0^1 \frac{\dd R}{R} \Psi_p(R) \Psi_q(R) = \delta(p-q) \, .
\end{align}

The normalizaton of the functions for the continuous spectrum can be found from their asymptotics. This method is described in detail in \cite{Landau:1965,Baz1969}.
Note that the asymptotics $Z_p(R\rightarrow0) \approx R^{ip} = e^{-ipx}$ imply
\begin{align}
\Psi_p(R\rightarrow0) \approx -f_p \Big[ \Re(c_p)\sin(px) + \Im(c_p) \cos(px) \Big] \, .
\end{align}
In order to extract the normalization factor $f_p$ we make use of these asymptotics. At $R\rightarrow 0$ one has
\begin{align}
\Psi_p \Psi_{k\approx p} \sim \frac12 |c_p|^2 f_p^2 \cos[(p-k)x] + \text{oscillating terms} \, .
\end{align}
At the same time, for plane waves one has $\varphi_p \varphi_k \sim 1/(2\pi)$. In order to have a similar normalization one finds
\begin{align}
f_p = \sqrt{\frac{2}{\pi|c_p|^2}} = \sqrt{\frac{2}{p\tanh(\pi p)}} \, .
\end{align}
The factor of $\sqrt{2}$ appears because
\begin{align}
\int\limits_{-\infty}^\infty \dd x \cos(px)\cos(kx) = \pi\delta(p+k) + \pi\delta(p-k) \, .
\end{align}

At the horizon one has
\begin{align}
\label{psi:horizon}
\Psi_p(1) = \sqrt{2p\coth(\pi p)} \, .
\end{align}
We arrive at a regular, real-valued expression for the function $F(s,R)$  \eq{FSR} with a proper normalization:
\begin{align}
\label{eq:regular}
F(s,R) = \int\limits_0^\infty \dd p \, \rho_p \, e^{-p^2 s} \Psi_p(R) .
\end{align}
This function is at the foundation of all our subsequent studies.

\subsection{Implementing $F(0,R) = R$}
\label{sec:f0r}
Inserting the boundary condition $F(0,R)=R$ into expression \eqref{eq:regular} gives
\begin{align}
1 = \int\limits_0^\infty \dd p \, \rho_p \, \frac{\Psi_p(R)}{R}
\end{align}
The orthogonality of the real regular solutions $\Psi_p(R)$ allows to invert this relation:
\begin{align}
\int\limits_0^1 \dd R \, \Psi_q(R) &= \int\limits_0^1 \dd R \int\limits_0^\infty \dd p \, \rho_p \, \Psi_q(R) \frac{\Psi_p(R)}{R} \\
&= \int\limits_0^\infty \dd p \rho_p \int\limits_0^1 \frac{\dd R}{R} \Psi_p(R) \Psi_q(R) \\
&= \int\limits_0^\infty \dd p \rho_p \delta(p-q) = \rho_q \, ,
\end{align}
Using this relation one can obtain the expression $\rho_p$ as follows.
Let us denote
\be
{\cal F}_p =  {}_3 F{}_2(1+ip,1+ip,ip;~ 2+ip, 1+2ip;~ 1 ) \, .
\ee
Then one has
\begin{align}\nonumber
d_p &= \int\limits_0^1 \dd R ~ Z_p(R) = \frac{{\cal F}_p}{1+ip}\, , \nonumber \\
\rho_p &= f_p \left[ \Re(c_p)\Im(d_p) - \Im(c_p) \Re(d_p) \right] \\
&= \frac{\sqrt{p\sinh(2\pi p)}}{\sqrt{2}\pi^{3/2}(1+p^2)} \Re \left[ \frac{i+p}{4^{ip}} \Gamma(ip)\Gamma\left(\tfrac12-ip\right) {\cal F}_p \right]  .\nonumber
\end{align}
It is possible to check that
\begin{align}
\int\limits_0^\infty \dd p \, \rho_p^2 = \frac 12 \, .
\end{align}

\subsection{Quasi-local approximation}
\label{sec:expansion}
Let us note that the function $F(s,R)$ can, at least formally, be expressed in the form of the following series
\begin{align}
F(s,R) &= e^{s\Box} R = \sum\limits_{n=0}^\infty \frac{s^n \Box^n}{n!} R \approx \sum\limits_{n=0}^N \frac{s^n \Box^n}{n!} R\\
&= R + s \partial_r f \partial_r R + \frac12 s^2 (\partial_r f \partial_r)^2 R + \mathcal{O}\left( s^{N+1} \right) \nonumber \, .
\end{align}
This representation by construction satisfies the boundary condition $F(0,R) = R$. One might expect that for a small value of the parameter $sR \ll 1$ it is sufficient to cut the series and to keep only a few first terms. As we will demonstrate below, this expectation is correct.
One can use these expressions to determine the influence of non-locality on the trace $T$ as well as the radial pressure $T^r_r$ and the Hawking flux.
Inserting this power series in \eqref{eq:trace-gf1} one finds the following series expansion for the trace:
\begin{align}
\begin{split}
&T = \sum\limits_{n=0}^\infty \frac{s^n}{n!} T_n(R) \, , \\
&T_n(R) = \frac{1}{48\pi} \left[ 2\Box^n R + \sum\limits_{p=0}^{n-1} (\Box^p R)(\Box^{n-p-1}R) \right] \, , \\
&T_\ins{Pol}(R) = \frac{R}{24\pi} \, .
\end{split}
\end{align}
In the above, $T_\ins{Pol}$ is the trace anomaly captured by the Polyakov action, and the terms $T_n$ with $n\ge 1$ contain non-local corrections from the ghost-free deformation of the Polyakov action.

\subsection{Results}

\subsubsection{Ghost-free contributions to the trace}
Having derived the explicit form of $F(s,R)$, we may now insert \eqref{eq:regular} into \eqref{eq:trace-gf1}. In order to study the contribution of GF modification to the trace anomaly we split it to the local term $T_\ins{Pol}$ coming from the Polyakov action and a GF correction $\Delta T$
\begin{align}
\begin{split}
T &= T_\ins{Pol} + \Delta T \hh
T{}_\ins{Pol} = \frac{1}{24\pi} R \, , \\
\Delta T &= \frac{1}{24\pi} \left[ F(s,R) - R \right] \\
&+\frac{s}{48\pi}\int\limits_0^1\dd\xi F[(1-\xi)s,R] F[\xi s,R] .
\end{split}
\end{align}
 The correction $\Delta T$ captures the non-local contributions to the trace anomaly. We evaluated $\Delta T$ using two approaces: (i) the continuous spectrum representation developed in Secs.~\ref{sec:spectral}--\ref{sec:f0r} as well as (ii) the approximate method detailed in Sec.~\ref{sec:expansion}.

\begin{figure*}[!hbt]
    \centering
    \subfloat{{ \includegraphics[width=0.47\textwidth]{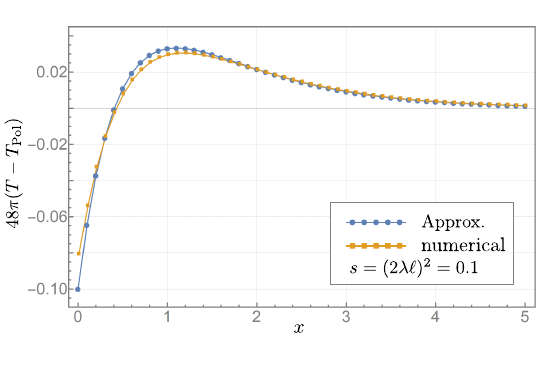} }}%
    \quad
    \subfloat{{ \includegraphics[width=0.47\textwidth]{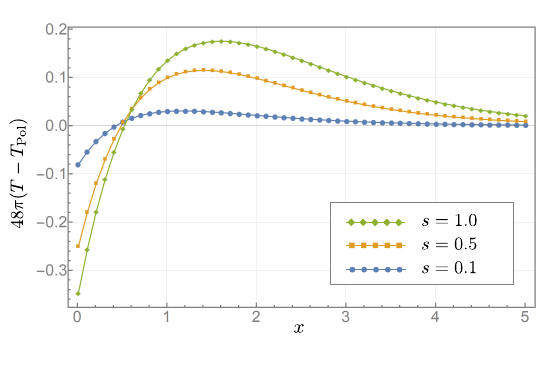} }}%
    \caption{Left: The non-local GF corrections to the trace plotted over the distance $x$ for a specific value of non-locality of $s=(2\lambda \ell)^2 = 0.1$, where $1/(2\lambda)$ is the characteristic length scale of the background black hole, and $\ell$ is the scale of non-locality. The numerical evaluation (labeled ``numerical'') agrees well with the small-$s$ expansion (labeled ``Approx.'' and performed to linear order in $s$). Right: Evaluated numerically non-local contributions to the trace anomaly are plotted over the distance $x$ for a few values of non-locality $s$.}
    \label{fig:trace}
\end{figure*}

 In Fig.~\ref{fig:trace} we presented the GF corrections to the trace of the stress-energy tensor computed using both approaches. One can see that $\Delta T$ is finite on the horizon and rapidly vanishes at infinity.
For small values of non-locality, $s = (2\lambda \ell)^2< 1$, both methods of computation agree within our resolution. We take it as an indication that our numerics work well. For values $s \gtrsim 1$  we cannot trust a series expansion anymore and therefore one can use only numerics to evaluate our exact representation.\footnote{In this case, however, note that a large non-locality $s$ \emph{improves} the numerical convergence.} in terms of hypergeometric functions. As depicted in Fig.~\ref{fig:trace}, the GF corrections grow for larger values of non-locality scale $\ell$ and decay slower at far distances.

\subsubsection{Ghost-free contribution to the black hole entropy}
In the case of dilaton gravity \eq{2BH}--\eq{rg} we can now compute the quantum corrections to the black hole entropy due to non-locality. As before, we split the entropy corrections into a well known local part, see \eqref{SPol}, as well as a non-local correction term $\Delta S$,
\begin{align}\label{SGF}
S_\ins{GF} &= S_\ins{Pol} + \Delta S .
\end{align}
The Polyakov contribution takes the form \cite{Myers:1994sg}
\be
S_\ins{Pol}=-{1\over 6}\phi=-{1\over 6}\lambda r_g=-{1\over 12}\ln{M\over\lambda} \, ,
\ee
where $\phi$ is the classical dilaton field.\footnote{Recall that the dilaton $\phi$ has nothing to to with the auxiliary scalar field $\varphi$, which we introduced to present  the Polyakov action in the local form \eq{eq:action}.} The non-local contribution $\Delta S$ to the black hole entropy \eq{SGF} reads
\begin{align}
\Delta S &= \frac{1}{12} \int\limits_0^s \dd \tilde{s} F(\tilde{s}, 1) \\
&= \frac{1}{12} \int\limits_0^\infty \dd p \, \rho_p \Psi_p(1) \frac{1-e^{-sp^2}}{p^2} .
\end{align}
It is a function of $\ell$ via $s = (2\lambda \ell)^2$.

Note that in this expression it is sufficient to use the values of the functions $F(s,R)$ and $\Psi_p(R)$  taken at the horizon, $R=1$. The considerations presented in Sec.~\ref{sec:spectral} guarantee that they are regular expressions. The multiplicative term involving $\ell$ has the following properties: In the limiting case $\ell\rightarrow 0$ it vanishes, so that one has $\Delta S = 0$, as it must be. Note that for arbitrary $\ell$ it is regular at $p=0$.

Unfortunately, an analytic evaluation of the integral is impossible, which is why we resort to numerical methods. As we have already demonstrated in the above, the numerical evaluation of $F(s,R)$ converges reliably. In this case, the integrand is a rapidly decreasing function of $p$, which greatly simplifies the numerics.

\begin{figure}[!htb]%
    \centering
    \includegraphics[width=0.47\textwidth]{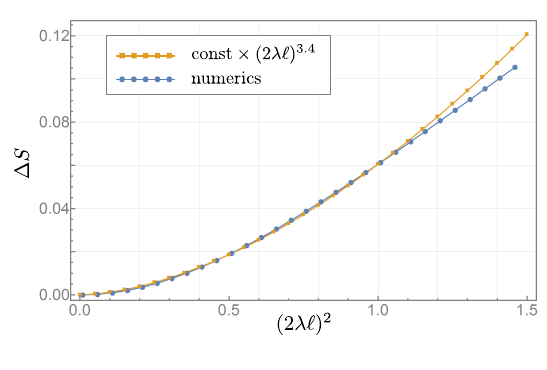}\\[-15pt]
    \caption{Ghost-free correction to the black hole entropy $\Delta S$ plotted as a function of non-locality $s = (2\lambda \ell)^2$.}
    \label{fig:deltaS}
\end{figure}

See Fig.~\ref{fig:deltaS} for a diagram of $\Delta S$ plotted as a function of non-locality $\ell$. In general, the corrections increase with a larger parameter of non-locality, and for $\ell=0$ they vanish, as expected. It is interesting to note that for small values, $s = (2\lambda \ell)^2 \lesssim 1$, the functional dependence on $\ell$ can be approximately captured by a power law,
\begin{align}
\Delta S(\ell < (2\lambda)^{-1}) \sim \text{const} \times s^{3.4} \, .
\end{align}
For larger values of non-locality, $s>1$, this approximation fails, but we are not aware of any closed form expression.

\section{Discussion}\label{sec:7}

Two dimensional dilaton gravity is often used for the modelling of properties of four-dimensional spacetimes. The reason is evident: they are much simpler and many problems can be solved exactly. Quantum theory of massless fields in 2D gravity is a well known example.
A two dimensional metric is conformally flat, so that the conformal invariance of such a theory reduces solving of the field equations to a similar problem in flat spacetime. The latter problem is technically much simpler. However, the calculation of local observables of a quantum conformal field requires renormalization, which breaks the conformal invariance. The conformal trace anomaly makes the quantum field feel the background. The response of the quantum average of the stress-energy tensor \SET\ can be obtained via the variation of the Polyakov effective action with respect to the two-dimensional metric tensor. In this paper we demonstrated how this effective stress-energy tensor depends on the choice of the quantum state and how these states are related to zero modes of the $\Box$-operator.

An interesting problem is how the effects of non-locality modify \SET. A natural way is to modify the kinetic term in the action for the quantum field by introducing the corresponding non-local form factor. For non-locality in the context of ghost-free $\mathrm{GF_N}$ theories this is equivalent to the substitution of $\Box \, \exp\left[(-\ell^2 \Box)^N\right]$ instead of the $\Box$-operator. Unfortunately, the calculation of the effective action for such a quantum theory becomes a very non-trivial problem. One reason is that this form factor breaks the original conformal invariance of the theory.

In the present paper we discuss another possible non-local modification of the Polyakov effective action. Namely, our starting point is a local action (\ref{a.4}), which is equivalent to the non-local Polyakov action, and which contains an auxiliary field $\varphi$. The corresponding modification implies the introduction of the form factor in the kinetic part of the auxiliary field. The corresponding non-local action takes the form (\ref{eq:action})--(\ref{P.1}). We obtained an expression for the non-local modification of the effective action, calculated the corresponding stress-energy tensor, and studied its dependence on the state described by zero modes of the $\Box$- operator. Our conclusion is that the main effect of non-locality is to modify the diagonal components of the stress-energy tensor, while the fluxes, described by non-diagonal components, remain unchanged.

In the application of these results to static two-dimensional black hole spacetimes this means that the back-reaction of the effective stress-energy tensor produced by the modification of the Polykov action changes the parameters of the black hole: its mass, surface gravity, and entropy. For a fixed background, however, the late-time Hawking flux of the energy at infinity remains unchanged.

To illustrate the effect of non-locality we considered a special metric which is a solution of the effective action for a two-dimensional string model. For this purpose we studied solutions of the eigenvalue problem $\Box \Psi_p=-p^2 \Psi_p$ for time-independent functions $\Psi_p$. In the considered metric this problem can be solved analytically. Moreover, we demonstrated that the corresponding spectrum is continuous and displayed the eigenfunctions explicitly. These results allowed us to calculate the stress-energy tensor for the effective action. We demonstrated that its deformation due to the presence of non-locality remains finite at the horizon. We also confirmed our general conclusion that the energy flux at infinity in a given fixed background is not affected by the presence of non-locality.

It would certainly be interesting to use the obtained results and developed tools for the study of back-reaction effects. In particular, it would be worthwhile to understand how these effects can change the structure of the black hole interior where non-local contributions to the stress-energy tensor become strong.

\section*{Acknowledgments}
J.B.\ is grateful for a Vanier Canada Graduate Scholarship administered by the Natural Sciences and Engineering Research Council of Canada as well as for the Golden Bell Jar Graduate Scholarship in Physics by the University of Alberta.
V.F.\ and A.Z.\ thank the Natural Sciences and Engineering Research Council of Canada and the Killam Trust for their financial support.

\appendix

\section{Useful formulas for two-dimensional static geometries}
\label{app:2d}
\label{app}

\subsection{General relations}

The geometry of a two-dimensional static spacetime is rather simple. We collect here useful formulae which are used in the main body of the paper.

Let us consider a two-dimensional metric $g_{\mu\nu}$ which admits a Killing vector $\xi^{\mu}$ such that
\be \n{A.1}
\xi_{(\mu;\nu)}=0\, .
\ee
We denote
\be
f= -\xi_{\mu}\xi^{\mu}\, .
\ee
We assume that the spacetime is asymptotically flat and normalize the Killing vector by the condition that at infinity $f=-1$. If this metric describes a 2D black hole, then $f=0$ at the event horizon.

In what follows, we shall focus on the exterior domain where $f\ge 0$. We denote by $\xi$ a one-form
\be \n{A.2}
\xi=\xi_{\mu} \dd x^{\mu}\,  .
\ee
The trace of equation (\ref{A.1}) implies that
\be \n{A.3}
\delta \xi=0\, ,
\ee
were $\delta=\star \dd \star$ denotes the exterior coderivative, $\dd$ is the exterior derivative, and $\star$ is the Hodge dual. The above relation implies that $\dd(\star\xi)=0$ and hence
\be \n{A.4}
\eta \equiv \star\xi = \dd r\, .
\ee
Here $r$ is a scalar function.

Since the relation $\xi \wedge \dd\xi=0$ is identically valid in 2D space, one has
\be \n{A.5}
\xi=-\beta \ \dd t\,  ,
\ee
where $t$ and $\beta$ are scalar functions. The minus sign in this relation is chosen for convenience.

Thus the Killing vector allows one to introduce special coordinates $(t,r)$. In these coordinates
\be \n{A.6}
g^{tr}=g^{\mu\nu} t_{,\mu} r_{\nu}=-\beta^{-1}\ (\star\xi ,\xi)=0\, .
\ee
One also has
\ba\n{A.7}
g^{tt}&=&g^{\mu\nu} t_{,\mu} t_{\nu}=-\beta^{-2}f\, ,\\
g^{rr}&=&g^{\mu\nu} r_{,\mu} r_{\nu}=(\star\xi,\star\xi)=f\, .
\ea
Thus the metric written in $(t,r)$ coordinates takes the form
\be\n{A.8}
\dd s^2=-{\beta^2\over f} \dd t^2+{1\over f}\dd r^2\, .
\ee
The relation $\xi^{\mu}(\xi^2)_{;\mu}=0$ implies that $f=f(r)$. The relation of $\xi^{(t; r)}=0$ for this metric gives
\be \n{A.9}
\beta' f-f'\beta=0\, ,
\ee
where $(...)'=\partial_r(...)$. This means that $\beta=\beta_0(t) f$. By redefinition of the coordinate $t$ the factor $\beta_0(t)$ can be put equal to 1. Thus the metric (\ref{A.8}) takes the form
\be\n{A.10}
\dd s^2=-f \dd t^2+{1\over f}\dd r^2=e^{2\sigma}\, (-\dd t^2+ \dd r_{*}^2)\, .
\ee
Here $\sigma={1\over 2}\ln f$ and $r_{*}$ is a tortoise coordinate. The following relation,
\be \n{A.11}
\Box \sigma =-R_{\mu\nu} {\xi^{\mu}\xi^{\nu}\over \xi^2} \, ,
\ee
is valid in any number of dimensions. In the 2D case, $R_{\mu\nu}={1\over 2}R g_{\mu\nu}$ and hence (\ref{A.11}) takes the form
\be \n{A.12}
\Box \sigma =-{1\over 2} R\, .
\ee
This means that a solution of the equation $\Box\varphi=R$ is
\be \n{A.13}
\varphi=-2\sigma+\chi=-\ln f +\chi\, ,
\ee
where $\chi$ is a `zero mode', that is, a solution of the homogeneous equation $\Box \chi=0$.

Let us demonstrate now that the functions $t$ and $r_*$ are zero mode solutions.
\ba \n{A.14}
\Box t&=&\delta (\dd t)=-\delta\left(\xi\over f\right)=-\star \dd \star\left(\xi\over f\right) \\
&=-& \star \left( \dd{\star\xi\over f}\right)=-{\delta\xi\over f}+{1\over f^2}\star(\star\xi\wedge \dd f)=0\, . \nonumber
\ea
In the last equation we used $\delta\xi=0$ and that both $\star\xi$ and $\dd f$ are proportional to $\dd r$.
Similarly one has
\be \n{A.15}
\Box r_*=\delta\left({dr\over f}\right)=\star\dd( {\star\dd r\over f} )=\star\dd\left( {\xi\over f} \right)=\star \dd^2 t=0\, .
\ee

Thus zero mode $\chi$ can be chosen as a linear combination (with constant coefficients) of two functions $t$ and $r_*$. One can also add a constant, but this trivial solution does not contribute to $T_{\mu\nu}$ and in what follows we shall ignore it for that reason. We shall use the functions
\be \n{A.16}
u=t-r_*\hh v=t+r_*\,  ,
\ee
instead of $t$ and $r_*$.  These functions are nothing but retarded and advanced time coordinates. For an eternal black hole, $u$ is regular at the past horizon, while $v$ is regular at the future horizon. They obey the equations
\be \n{A.17}
\Box u=\Box v=0\, .
\ee

In the next subsection we show how zero mode solutions are related to the choice of the state in the theory. To finish this section we add two useful expressions for the mass function and the surface gravity of a two dimensional black hole.

Suppose there exists a conserved symmetric tensor $T_{\mu\nu}$, $T_{\mu\nu}{}^{;\nu}=0$.
Let us denote $J_{\mu}=T_{\mu\nu}\xi^{\nu}$ a Killing current connected with this tensor. Then one has
\be
\dd(\star J)=\star\delta(J)=0\, .
\ee
This relation implies that the form $\star J$ is closed and there exist such a mass function $m$ that
\be
\dd m=-\star J\, .
\ee
For a stationary tensor $T_{\mu\nu}$ in the coordinates (\ref{A.10}) one has
\be
m=-\int \dd r \,T_0^0 \, .
\ee
This relation allows one to obtain a contribution to the mass of a black black by the effective stress-energy tensor of a test field calculated on the black-hole background.

One can also prove the following useful formula for the surface gravity of a two-dimensional black hole \cite{Frolov:1992xx},
\be
\kappa={1\over 2} \int\limits_\Sigma R \xi^{\mu} \dd\Sigma_{\mu}\, .
\ee
Here $\Sigma$ is a one-dimensional surface (line) between the horizon and infinity, and $\dd\Sigma_{\mu}$ is the corresponding surface element. In the coordinates (\ref{A.10}) this formula takes the form
\be
\kappa={1\over 2} \int\limits_{r_g}^{\infty}  R\, \dd r={1\over 2}f'|_{r_g}\, .
\ee

\subsection{The stress-energy tensor}

We demonstrate now that different choices of zero mode functions $\chi$ in a solution for the auxiliary field $\varphi$, see Eq.~(\ref{A.15}), result in a special form of the effective stress-energy tensor related to a special choice of the corresponding quantum state.

\subsubsection{Boulware vacuum}

Let us put $\varphi=-\ln f$. The calculations give
\be \n{A.18}
b^{-1} t^{\mu}_{\nu}=\mbox{diag} \left(\frac{f'^2}{2 f}-2f'',-\frac{f'^2}{2 f}\right)\hhh t^{\mu}_{\mu}=-2f''=2R\, .
\ee
This expression vanishes at $\mathscr{I}^+$ and $\mathscr{I}^-$ and is singular at both future and past horizons.
Hence it correctly reproduces the quantum average of the stress-energy tensor in the Boulware vacuum state.

\subsubsection{Hartle--Hawking vacuum}

Let us put $\varphi=-\ln f+k r_*$. One has
\be \n{A.19}
t^{\mu}_{\nu}=b^{-1} T^{\mu}_{\nu}=\mbox{diag}\left({f'^2-k^2\over 2 f}-2f'',-{f'^2-k^2\over 2 f}\right) \, .
\ee
For a general value of $k$ this stress-energy tensor diverges at the horizons. However, it remains finite for a special case
$k=f'|_{r_g}=2\kappa$, where $\kappa$ is the surface gravity. For this case at infinity
\be
 t^{\mu}_{\nu}\sim\mbox{diag} (-2\kappa^2 ,2\kappa^2)\, .
\ee
The corresponding state in this case is the Hartle--Hawking vacuum.

\subsubsection{Unruh vacuum}

Let us put $\varphi=-\ln f+\kappa u$. Then in $(t,r)$ coordinates one has
\ba \n{A.20}
&&t_t^{\ t}=-2f''+{1\over 2f}(f'-2\kappa^2)\hhh t_t^{\ r}=-\kappa^2 \, , \nonumber\\
&& t_r^{\ t}={\kappa^2\over f^2}\hhh t_r^{\ r}=-{1\over 2f}(f'-2\kappa^2)\, .
\ea
Let us denote by $U_{\mu}=(-f,1)$ a null vector which is regular at infinity. Then, at large $r$, one has $t_\mu{}^\nu\sim \kappa^2 U_{\mu} U^{\nu}$. Hence the corresponding stress-energy tensor describes an out-going flux of null fluid (radiation) at $\mathscr{I}^+$.

Let us demonstrate now that the stress-energy tensor \eqref{A.20} is regular at the future event horizon. To demonstrate this we write our metric in advanced time coordinates $v=t+r_*$ that are regular at the future horizon
\begin{align}
\dd s^2=-\dd v^2+2 \dd v \dd r\, .
\end{align}
The calculations give
\ba
&& t_{vv}=2f f''-{1\over 2}f'^2+\kappa^2\, ,\nonumber\\
&&t_{vr}=t_{tv}=-4f''+{1\over 2f}(f'^2-4\kappa^2)\, ,\\
&& t_{rr}=2{f''\over f}-{1\over f^2}(f'^2-4\kappa^2)\, .\nonumber
\ea
Near the horizon one has
\be
f=2\kappa (r-r_g)+{1\over 2}f_2  (r-r_g)^2+{1\over 6}f_3  (r-r_g)^3+\ldots\, .
\ee
One has
\ba
&&t_{v v}= -\kappa^2+O(r-r_g)\hhh
t_{vr}=-f_2+O(r-r_g) \, , \nonumber\\
&&t_{rr}=f_3/\kappa+O(r-r_g)\, .
\ea
These relations imply regularity of $t_{\mu\nu}$ at the future horizon. Hence, this stress-energy tensor possesses the proper boundary condition required for the Unruh vacuum state. One can also see that negative energy flux through the horizon, $t_{vv}|_{r_g}=-\kappa^2$, is equal (with a minus sign) to the outgoing energy flux at $\mathscr{I}^+$, $t_{uu}|_{J^+}=\kappa^2$.



%

\end{document}